%% file: MLCG.tex
\documentclass[sn-mathphys,Numbered]{sn-jnl}
\usepackage{color}
\usepackage{graphicx}%
\usepackage{multirow}%
\usepackage{amsmath,amssymb,amsfonts}%
\usepackage{amsthm}%
\usepackage{mathrsfs}%
\usepackage[title]{appendix}%
\usepackage{xcolor}%
\usepackage{textcomp}%
\usepackage{manyfoot}%
\usepackage{booktabs}%
\usepackage{algorithm}%
\usepackage{algorithmicx}%
\usepackage{algpseudocode}%
\usepackage{listings}%

\begin{document}

\title[Navigating protein landscapes with a transferable coarse-grained model]{Navigating protein landscapes with a machine-learned transferable coarse-grained model}

\author[1,2,3]{\fnm{Nicholas E.}\sur{Charron}}
\author[1]{\fnm{Felix}\sur{Musil}} 
\author[1]{\fnm{Andrea}\sur{Guljas}} 
\author[4]{\fnm{Yaoyi}\sur{Chen}}
\author[1]{\fnm{Klara}\sur{Bonneau}}
\author[1]{\fnm{Aldo S.}\sur{Pasos-Trejo}}
\author[1,4]{\fnm{Jacopo}\sur{Venturin}}
\author[1]{\fnm{Daria}\sur{Gusew}}
\author[1,2,5]{\fnm{Iryna}\sur{Zaporozhets}}
\author[4]{\fnm{Andreas}\sur{Kr\"amer}}
\author[1]{\fnm{Clark}\sur{Templeton}}
\author[4]{\fnm{Atharva}\sur{Kelkar}}
\author[4]{\fnm{Aleksander E. P.}\sur{Durumeric}}
\author[6]{\fnm{Simon}\sur{Olsson}}
\author[7,8]{\fnm{Adrià}\sur{Pérez}}
\author[7,8]{\fnm{Maciej}\sur{Majewski}}
\author[9]{\fnm{Brooke E.}\sur{Husic}}
\author[10]{\fnm{Ankit}\sur{Patel}}
\author*[7,8,11]{\fnm{Gianni}\sur{De Fabritiis}}\email{g.defabritiis@gmail.com}
\author*[1,4,5,12]{\fnm{Frank}\sur{No\'e}}\email{franknoe@microsoft.com}
\author*[1,2,3,5]{\fnm{Cecilia}\sur{Clementi}}\email{cecilia.clementi@fu-berlin.de}

\affil[1]{\orgdiv{Department of Physics}, \orgname{Freie Universit\"at Berlin}, \orgaddress{\street{Arnimallee 12}, \city{Berlin}, \postcode{14195}, \country{Germany}}}

\affil[2]{\orgdiv{Department of Physics}, \orgname{Rice University}, \orgaddress{\street{6100 Main Street}, \city{Houston}, \postcode{77005}, \state{Texas}, \country{USA}}}

\affil[3]{\orgdiv{Center for Theoretical Biological Physics}, \orgname{Rice University}, \orgaddress{\street{6100 Main Street}, \city{Houston}, \postcode{77005}, \state{Texas}, \country{USA}}}

\affil[4]{\orgdiv{Department of Mathematics and Computer Science}, \orgname{Freie Universität Berlin}, \orgaddress{\street{Arnimallee 12}, \city{Berlin}, \postcode{14195}, \country{Germany}}}

\affil[5]{\orgdiv{Department of Chemistry}, \orgname{Rice University}, \orgaddress{\street{6100 Main Street}, \city{Houston}, 
\postcode{77005}, \state{Texas}, \country{USA}}}

\affil[6]{\orgdiv{Department of Computer Science and Engineering}, \orgname{Chalmers University of Technology}, \orgaddress{\street{Chalmersplatsen 4}, \city{Gothenburg}, \postcode{412 96}, \country{Sweden}}}

\affil[7]{\orgdiv{Computational Science Laboratory}, \orgname{Universitat Pompeu Fabra, Barcelona Biomedical Research Park (PRBB)}, \orgaddress{\street{Carrer Dr. Aiguader 88}, \city{Barcelona}, \postcode{08003}, \country{Spain}}}

\affil[8]{\orgname{Acellera Labs}, \orgaddress{\street{Doctor Trueta 183}, \city{Barcelona}, \postcode{08005}, \country{Spain}}}

\affil[9]{\orgdiv{Lewis Sigler Institute for Integrative Genomics}, \orgname{Princeton University}, \orgaddress{\street{330 Alexander Street}, \city{Princeton}, \postcode{08540}, \state{NJ}, \country{USA}}}

\affil[10]{\orgdiv{Department of Neuroscience}, \orgname{Baylor College of Medicine}, \orgaddress{\street{1 Baylor Plaza}, \city{Houston}, \postcode{77030}, \state{Texas}, \country{USA}}}

\affil[11]{\orgname{Instituci\'o Catalana de Recerca i Estudis Avanc¸ats (ICREA)}, \orgaddress{\street{Passeig Lluis Companys 23}, \city{Barcelona}, \postcode{08010},  \country{Spain}}}

\affil[12]{\orgdiv{AI4Science}, \orgname{Microsoft Research}, \orgaddress{\street{Karl-Liebknecht Str. 32}, \city{Berlin}, \postcode{10178}, \country{Germany}}}

\abstract{
The most popular and universally predictive protein simulation models employ all-atom molecular dynamics (MD), but they come at extreme computational cost. The development of a universal, computationally efficient coarse-grained (CG) model with similar prediction performance has been a long-standing challenge.
By combining recent deep learning methods with a large and diverse training set of all-atom protein simulations, we here develop a bottom-up 
CG force field with chemical transferability, which can be used for extrapolative molecular dynamics on new sequences not used during model parametrization. We demonstrate that the model successfully predicts folded structures, intermediates, metastable folded and unfolded basins, and the fluctuations of intrinsically disordered proteins while it is several orders of magnitude faster than an all-atom model.
This showcases the feasibility of a universal and computationally efficient machine-learned CG model for proteins.
}

\maketitle

\section{Introduction}
Large molecular systems are inherently multiscale and therefore choosing a suitable time- and length-scale in a computational model is challenging. A model must be fine-grained enough to make accurate predictions, but unnecessary detail should be avoided to increase computational efficiency and facilitate the understanding of how the macroscopic observables arise from the interaction of the microscopic degrees of freedom. For the important task of modeling macromolecular changes, such as protein folding or protein-ligand binding, and predicting their thermodynamic properties, the most successful and widely used simulation approach is molecular dynamics (MD) with all-atom resolution~\cite{freddolino2008ten,shaw2010atomic,voelz2010molecular,larsen_fast_folder_2011,zhao2013mature,plattner2017complete,robustelli2022molecular}. However, all-atom MD comes at extreme computational costs and it requires significant effort to post-process and analyze the data~\cite{ChoderaNoe_COSB14_MSMs,HusicPande_MSMReview,plattner2017complete}. It is still unclear if there is a computationally efficient coarse-grained scale that lends itself to a general simulation model. On the other hand, deep-learning methods have been wildly successful at predicting protein structure and function by reasoning over large-scale genomic and structure datasets~\cite{AlphaFold2021,Lin2023_ESM2}, but often do not tie into a physical level of understanding. In this manuscript, we provide evidence that using deep learning technologies a universal coarse-grained protein force field can be developed that can predict protein structures, structure transitions, and folding mechanisms similar to its all-atom counterparts while being orders of magnitude faster.  

Over the last fifty years,  significant developments in hardware, software, and theory have advanced the simulation of macromolecules from proof of principle to \textit{in silico} study of protein folding and dynamics~\cite{mccammon1977dynamics,garcia1992large,ferrara2000thermodynamics,freddolino2008ten,shaw2010atomic,voelz2010molecular,larsen_fast_folder_2011,zhao2013mature,plattner2017complete,robustelli2022molecular}. 
Most MD simulation studies employ 
atomistic force fields 
fitted on a combination of quantum chemical calculations and experimental data. The latest atomistic force fields have been shown to be qualitatively accurate for processes on nanosecond to millisecond timescales, while quantitatively their predictions are typically within a few kcal/mol compared to experiment 
\cite{Lindorff-Larsen_Piana_Palmo_Maragakis_Klepeis_Dror_Shaw_2010,plattner2017complete,CasanovasParrinello_Dissociation}.
Recently introduced machine-learned force fields~\cite{UnkeEtAl_Gems,Musaelian_AllegroBiosim,SmithIsayevRoitberg_ANI} may capture the quantum-mechanical interactions between nuclei in the Born-Oppenheimer approximation even more accurately, but also come at higher computational cost compared to empirical all-atom force fields.

Despite the impressive results obtained with all-atom MD simulations, ever since the first protein simulations~\cite{levitt1975computer} the community has striven to develop universal coarse-grained (CG) macromolecular models that are computationally more efficient and simpler to analyze. 

Statistical mechanical models of protein folding, such as the energy landscape theory~\cite{onuchic1997theory}, compounded with the results from decades of analysis of atomistic simulations~\cite{Noe2017}, 
suggest that not every atom is \emph{per se} important in the definition of a protein's free energy landscape, but that a small(er) number of collective variables are sufficient to realistically describe it.   
Structure-based models~\cite{clementi2000topological, de2022smog} build a force field around the known native structure of a protein 
to explore its free energy landscape.
MARTINI~\cite{Martini3} is a widely-used CG force field that has been very successful in modeling molecular interactions, such as membrane structure formation and protein interactions, but does not typically predict intramolecular protein dynamics such as folding and other large conformational changes.
In parallel, significant efforts have been devoted to the development of CG force fields for protein folding and conformational dynamics such as UNRES~\cite{Liwo_Sieradzan_Lipska_Czaplewski_Joung_Zmudzinska_Halabis_Oldziej_2019} or AWSEM~\cite{Davtyan_Schafer_Zheng_Clementi_Wolynes_Papoian_2012}, which have been used to predict folded structures but often do not capture alternative metastable states, or free energy minima, that appear if all-atom force fields are used instead. 

Bottom-up CG force fields 
\cite{joshi2021review,giulini2021system,Jin_Pak_Durumeric_Loose_Voth_2022,noid2023perspective},
which are typically fit to match the equilibrium distribution of an all-atom model, could in principle reach the accuracy and predictiveness of the latter, but have not, as of yet, realized their full potential.
The main hindrance to the development of an accurate biomolecular CG model is that even if classical atomistic force fields usually model non-bonded interactions as a sum of two-body terms, CG force fields require multi-body functional forms to realistically represent protein thermodynamics and implicit solvation effects~\cite{Wang_Charron_Husic_Olsson_Noe_Clementi_2021}. 
By leveraging recent developments in deep learning it has become possible to machine-learn such many-body CG force fields using neural networks 
\cite{lemke2017neural,Wang_2019,Husic_Charron_Lemm_2020,Wang_Charron_Husic_Olsson_Noe_Clementi_2021,Ding_Zhang_2022,Majewski_Perez_Tholke_Doerr_Charron_Giorgino_Husic_Clementi_Noe_DeFabritiis_2022,Koehler_Chen_Kraemer_Clementi_Noe_2023,chennakesavalu2023ensuring,kramer_noise_opt_2023,wellawatte2023neural,airas2023transferable,ArtsEtAl_TwoForOne}.
In particular using the variational force-matching approach \cite{Izvekov_Voth_2005,izvekov2005multiscale,izvekov2006multiscale,Zhou_Thorpe_Izvekov_Voth_2007,Noid_Chu_Ayton_Krishna_Izvekov_Voth_Das_Andersen_2008,Thorpe_Zhou_Voth_2008,Jr_Lu_Voth_2010,Thorpe_Goldenberg_Voth_2011}, such machine-learned CG force fields have been shown to accurately represent the coarse-grained all-atom distributions of single proteins 
\cite{Wang_2019,Husic_Charron_Lemm_2020,Wang_Charron_Husic_Olsson_Noe_Clementi_2021,Chen_Kramer_Charron_Husic_Clementi_Noe_2021,kramer_noise_opt_2023,Koehler_Chen_Kraemer_Clementi_Noe_2023} and of multiple proteins~\cite{Majewski_Perez_Tholke_Doerr_Charron_Giorgino_Husic_Clementi_Noe_DeFabritiis_2022}. 
However, a transferable CG force field that could be considered quantitative, predictive, and as reliable as a modern atomistic force field is still missing~\cite{Durumeric_Charron_Templeton_Musil_Bonneau_Pasos-Trejo_Chen_Kelkar_Noe_Clementi_2023}.

Here we propose a CG model that is truly transferable in sequence space: we learn the model parameters on a dataset of atomistic simulations for a set of proteins and then use it to successfully simulate the folding/unfolding dynamics of different proteins, with sequences never seen in the learning stage, and with low ($<$ 50\%) sequence identity with the proteins used in the training or validation sets. The CG model is orders of magnitude faster than all-atom MD simulations, provides consistent predictions for systems where all-atom MD simulations are affordable, and is consistent with experimental data for larger proteins where converged all-atom simulations are not available. This result indicates that the neural network ``learns'' to represent the effective physical interactions between the CG degrees of freedom and provides strong support for the hypothesis that using deep learning methods, a universal CG model for realistic and predictive protein simulations at low computational cost is within reach.
\section{Results}
\begin{figure}[h] \includegraphics[width=\textwidth]{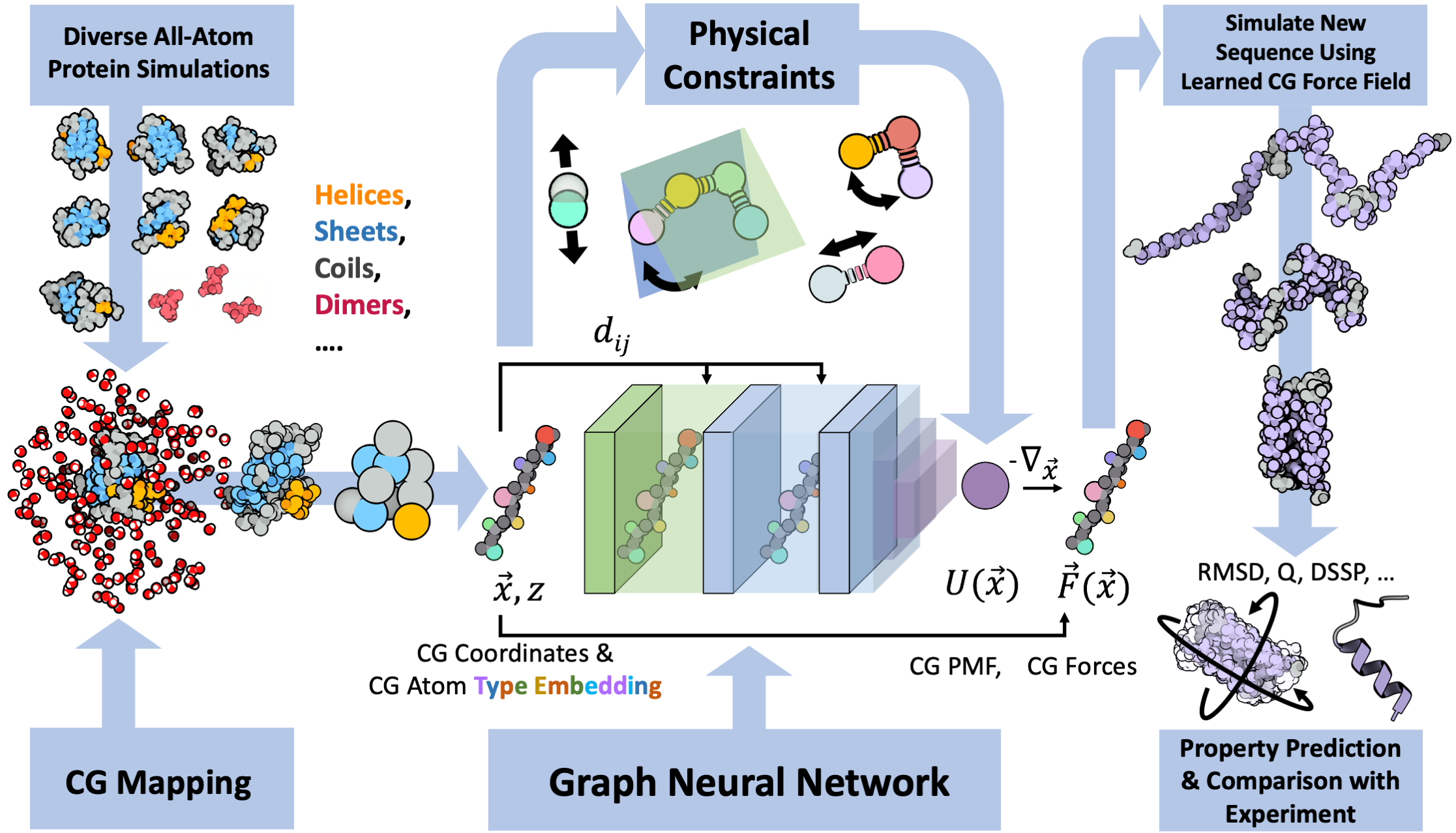}
    \caption{Conceptual illustration of the pipeline for building and testing a transferable, bottom-up, machine-learned, CG protein force field from a diverse dataset of all-atom simulations, a chosen CG resolution, and a set of basic physical prior energy terms (bonds, angles, dihedrals, and purely repulsive interactions).}
\label{pipeline}
\end{figure}

We generated a dataset of all-atom simulations of 50 CATH domains~\cite{cath}, representing small proteins with diverse folds as well as $\approx$1200 dimers of mono and dipeptides in explicit solvent. We stored all instantaneous forces on the protein atoms, performed force aggregation on a CG backbone representation of the proteins~\cite{kramer_noise_opt_2023}, and trained a CG force field CGSchNet~\cite{Husic_Charron_Lemm_2020} which combines a deep graph neural network (GNN) with physically motivated prior terms. We conduct a series of extensive Langevin and parallel-tempering simulations of the learned model on proteins of various sizes, secondary, and tertiary structures in order to demonstrate its capabilities and limitations. Wherever feasible we have also performed extensive all-atom MD simulations for the test systems and promoted their convergence with Markov state modeling \cite{ChoderaNoe_COSB14_MSMs,HusicPande_MSMReview} for comparison. See Fig.~\ref{pipeline}, Methods, and Supplementary Information (SI) for details.
\subsection{Conformational landscape of peptides and small proteins}
\begin{figure}[h]
    \centering
\includegraphics[width=\textwidth]{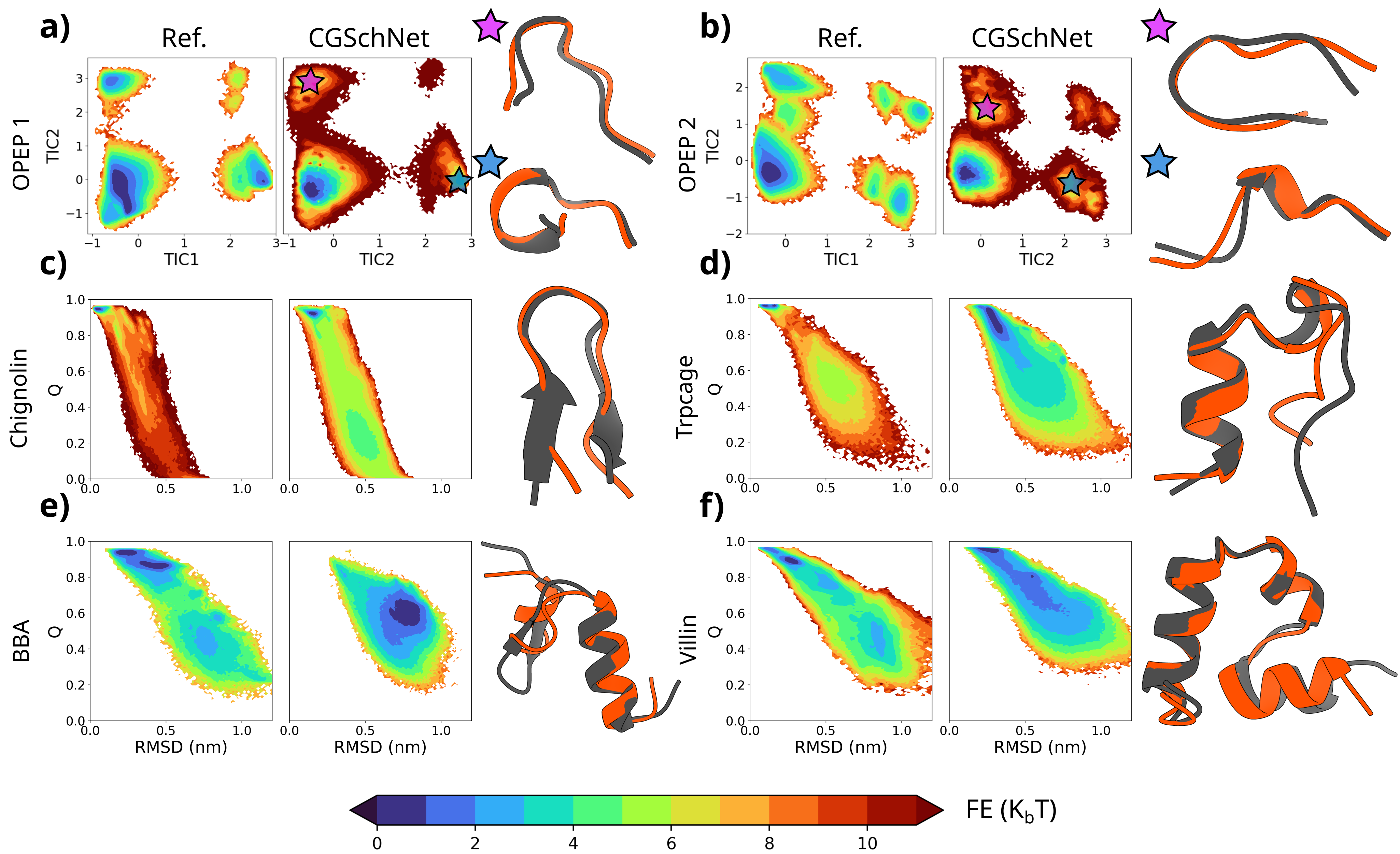}
    \caption{CGSchNet performance on peptides and small proteins: a) 8-residue peptide DYGCSIHP, b) 8-residue peptide SLEAGGRG, c) CLN025 (2RVD), d) Trpcage (2JOF), e) BBA (1FME), and f) villin (1YRF). These sequences are completely absent from the training and validation sets.
    In each subfigure, the 2D free energy surface of the CG model (CGSchNet) is shown in comparison with reference atomistic simulations at 300K. The free energy is shown as a function of the first two TICA components~\cite{Perez-Hernandez_2013} for the two 8-peptides, and as a function of the fraction of native contacts, $Q$, and the carbon alpha RMSD to the native state for the four small proteins. Comparisons of structures from the most folded-like basin (or labeled basins for the 8-peptides) for CGSchNet (orange) and atomistic (grey) models are also shown. The free energy landscapes for the CG model are obtained through WHAM-reweighted parallel-tempering simulations. See SI for more simulation details.}
\label{small_protein_results}
\end{figure}

To assess the ability of our approach to learning a \textit{transferable} CG force field, we first test how it reproduces the folding/unfolding free energy landscape of all-atom MD simulations for a set of 8-peptides and small fast-folding proteins: the 025 mutant of chignolin~\cite{Honda_Akiba_Kato_Sawada_Sekijima_Ishimura_Ooishi_Watanabe_Odahara_Harata_2008} (PDB code 2RVD, Fig.~\ref{small_protein_results}c), Trpcage~\cite{Barua_Lin_Williams_Kummler_Neidigh_Andersen_2008} (2JOF, Fig.~\ref{small_protein_results}d),
BBA~\cite{Sarisky_Mayo_2001} (1FME, Fig.~\ref{small_protein_results}e), and Villin headpiece~\cite{Chiu_Kubelka_Herbst-Irmer_Eaton_Hofrichter_Davies_2005} (1YRF, Fig.~\ref{small_protein_results}f). Free energy surfaces of the CG model were obtained through parallel-tempering (PT) simulations, a commonly employed enhanced simulation technique, to ensure converged sampling of the equilibrium distribution (see SI for details).
For these small proteins, we have obtained converged folding/unfolding reference landscapes from atomistic simulations for comparison, a task that quickly becomes computationally unfeasible for larger proteins.
None of these proteins have sequence similarity $> 50$\% with any stretch of sequence from the proteins in the training or validation datasets (see SI for details).

The free energy landscapes of the two 8-peptides match the atomistic references closely (Fig. \ref{small_protein_results}a-b). These peptides are mostly disordered with little, if any, secondary structure, and their landscape is mostly determined by the torsional dynamics contained in the prior energy term of the model, whereas the machine-learned multi-body terms have a small effect on the result.
The situation is very different for the four small fast-folding proteins (Fig. \ref{small_protein_results}c-f): the ability to predict the configurational landscape of these proteins is a learned property of the network and not included in the prior energy term. Control simulations with only the prior energy term only visit the unfolded state (SI Fig.~S5). 

For all four fast-folding proteins, the CG model predicts metastable folding and unfolding transitions, \textit{i.e.}, both folded and unfolded states have a free energy minimum. The folded states are predicted with a fraction of native contacts $Q$ close to $1$, and low RMSD values, closely resembling the correct native state (Fig. \ref{small_protein_results}).
Interestingly, for CLN025, the model is also able to stabilize the same misfolded state with misaligned TYR1 and TYR2 residues as found in the reference atomistic simulations (see  Fig. \ref{fig:comparison_landscapes} below).

For 3 of the 4 proteins in Fig. \ref{small_protein_results} the free energy basin containing the native state is the global minimum, while for protein BBA it is a local minimum, indicating that all proteins are able to fold/unfold correctly. 
However, the relative free energy difference between the folded and unfolded states does not exactly match those from the reference atomistic free energy surfaces. 
For Chignolin, Trpcage and Villin the model performs much better than for BBA. BBA contains both helical and anti-parallel beta-sheet motifs, and the difficulty of correctly stabilizing its folded state with CG models has been reported in previous works~\cite{Majewski_Perez_Tholke_Doerr_Charron_Giorgino_Husic_Clementi_Noe_DeFabritiis_2022, Liwo_Czaplewski_Pillardy_Scheraga_2001} using bottom-up or partially bottom-up approaches, often with concern that the stabilization of beta sheets requires specific higher body-order terms in CG models~\cite{Pillardy_Czaplewski_Liwo_Wedemeyer_Lee_Ripoll_Arlukowicz_Oldziej_Arnautova_Scheraga_2001}.
While Fig. \ref{small_protein_results} focuses on the folding/unfolding pathway as shown in the collective variables RMSD and fraction of native contacts, it is interesting that our learned CG model recapitulates many more characteristics and metastable intermediate structures from the all-atom model, as will be discussed below.

\subsection{Computational efficiency of CG simulations}
We compare the computational efficiency of our CGSchNet model and all-atom simulation. On one hand, conducting a single simulation timestep with the neural network model in PyTorch takes more wall-clock time than a single all-atom timestep in OpenMM using the same GPU (Table S11). That may seem counterintuitive as the CG model has many fewer degrees of freedom. However the all-atom potential is much simpler and MD simulation codes such as OpenMM are highly optimized for speed, whereas deep learning codes such as PyTorch are optimized for parallel processing. Despite this ``slowdown'' per simulation step, the CG simulations are overall much more efficient, because one CG simulation step corresponds to a much longer physical time equivalent. Two factors compound here: limiting the CG representation to fewer beads with effective interactions allows the physical time step length to be larger than in an all-atom simulation. 
Additionally, the CG landscapes appear overall smoother than their atomistic counterparts, with lower energy barriers separating folded and unfolded states, as expected from CG models as a result of averaging over part of the degrees of freedom. 

To quantify the effective speedup we estimate how much wall-clock time is needed, in either the CG or the all-atom model, to get one uncorrelated sample for the four small folders in Fig. \ref{small_protein_results}. This is measured in terms of the slowest relaxation timescale of a Markov state model \cite{Prinz_2011}, which is typically due to the folding/unfolding transition for small folders. Depending on the protein, the CG model is between a factor of 30 and 500 faster than all-atom simulations on the same GPU (see Table S11). 

There is an additional and very substantial efficiency gain in terms of throughput: while OpenMM is designed for one simulation per GPU and the throughput does not significantly increase from running multiple simulations on one GPU, batch simulation is trivial with the PyTorch code. For the proteins shown here, this results in an additional $5\times$ to $20\times$ increase in simulation throughput on top of the speedup discussed above. We note that in principle a substantial throughput increase might also be possible for all-atom simulation codes if their codes were designed for batch simulation.

\subsection{Extrapolation on larger proteins}
\begin{figure}[H]
    \centering   \includegraphics[width=\textwidth]{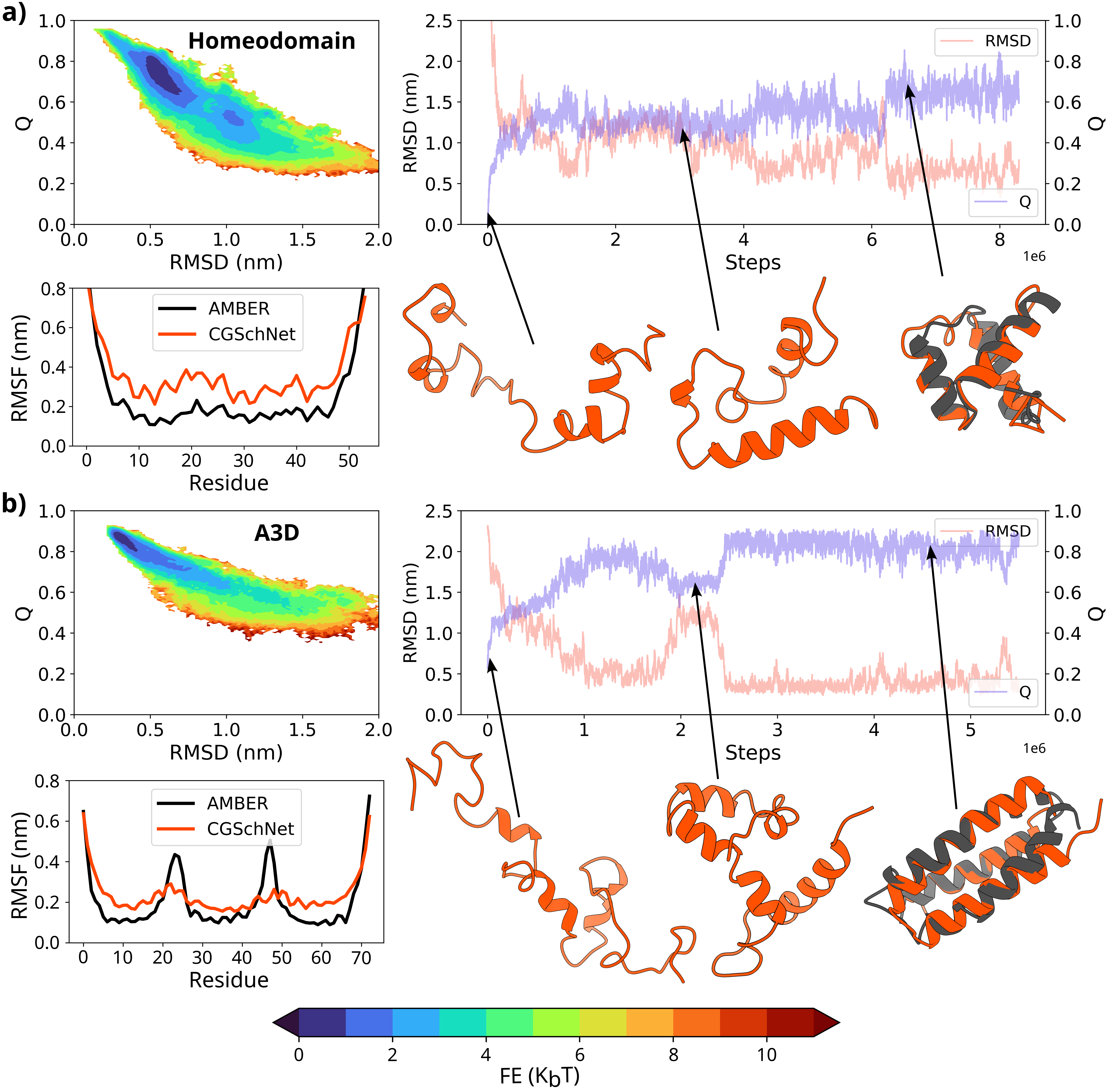}
\caption{Extrapolative performance of CGSchNet on two large proteins withheld from the training and validation sets. For each protein, the carbon-alpha root mean square fluctuations of the folded state to the crystal structure are shown and compared to the ones of the reference (AMBER) all-atom simulations. An exploration of the free energy surface as a function of the fraction of native contacts, $Q$, and the carbon alpha RMSD to the native state, obtained through PT simulations is represented. Furthermore, a folding trajectory starting from a completely elongated structure is shown, together with structures illustrating folding and a comparison with the crystal structure (gray). a) 54-residue Homeodomain (PDB code 1ENH), b) 73-residue alpha3D (PDB code 2A3D). For the computation of the RMSF of the CGSchNet model, a window of a trajectory staying folded for more than 1 million MD steps was chosen. More details can be found in the SI.
\label{large_protein_results}}
\end{figure}

To assess the ability of our CG model to fold and maintain folded states of larger systems with more complicated structures, we consider the following two proteins: 54-residue Engrailed Homeodomain (1ENH)~\cite{Clarke_Kissinger_Desjarlais_Gilliland_Pabo_1994}, and 73-residue de novo designed protein alpha3D (2A3D)~\cite{Walsh_Cheng_Bryson_Roder_DeGrado_1999} (Fig. \ref{large_protein_results}).

The size of these proteins does not allow atomistic simulations to sample the folding/unfolding transitions in a reasonable time frame.
Instead, we performed simulations with the reference atomistic force field in the folded state and compared the folded state dynamics.
On the other hand, the full free energy landscape can be easily explored by the CG model.
We perform MD simulations with our CG model and we define the free energy minimum with the highest fraction of native contacts, $Q$, as the folded state.
We then compare the C$_\alpha$ root mean square fluctuations (RMSF) within the CG native state free energy minimum with respect to the reference folded state all-atom simulations.

In the case of Homeodomain, the CG model stabilizes a state very close to the reference native structure, with higher relative flexibility at the N and C termini comparable with the all-atom simulations. However, there is a slight increase in flexibility along the entire sequence (Fig. \ref{large_protein_results}a).
For alpha3D, the CG model correctly stabilizes a single metastable state that is very close to the reference native structure. The increased flexibility found in the reference simulations at the termini and the turns between each helical bundle is also reproduced (Fig. \ref{large_protein_results}b). As in the case of Homeodomain, the CG model tends to be more flexible overall than the all-atom model.

These results show that the transferable machine-learned CG model can extrapolate to larger unseen proteins, stabilizing the correct native states and backbone fluctuation spectra associated with them.

\begin{figure}[H]
    \centering   \includegraphics[width=0.86\textwidth,keepaspectratio]{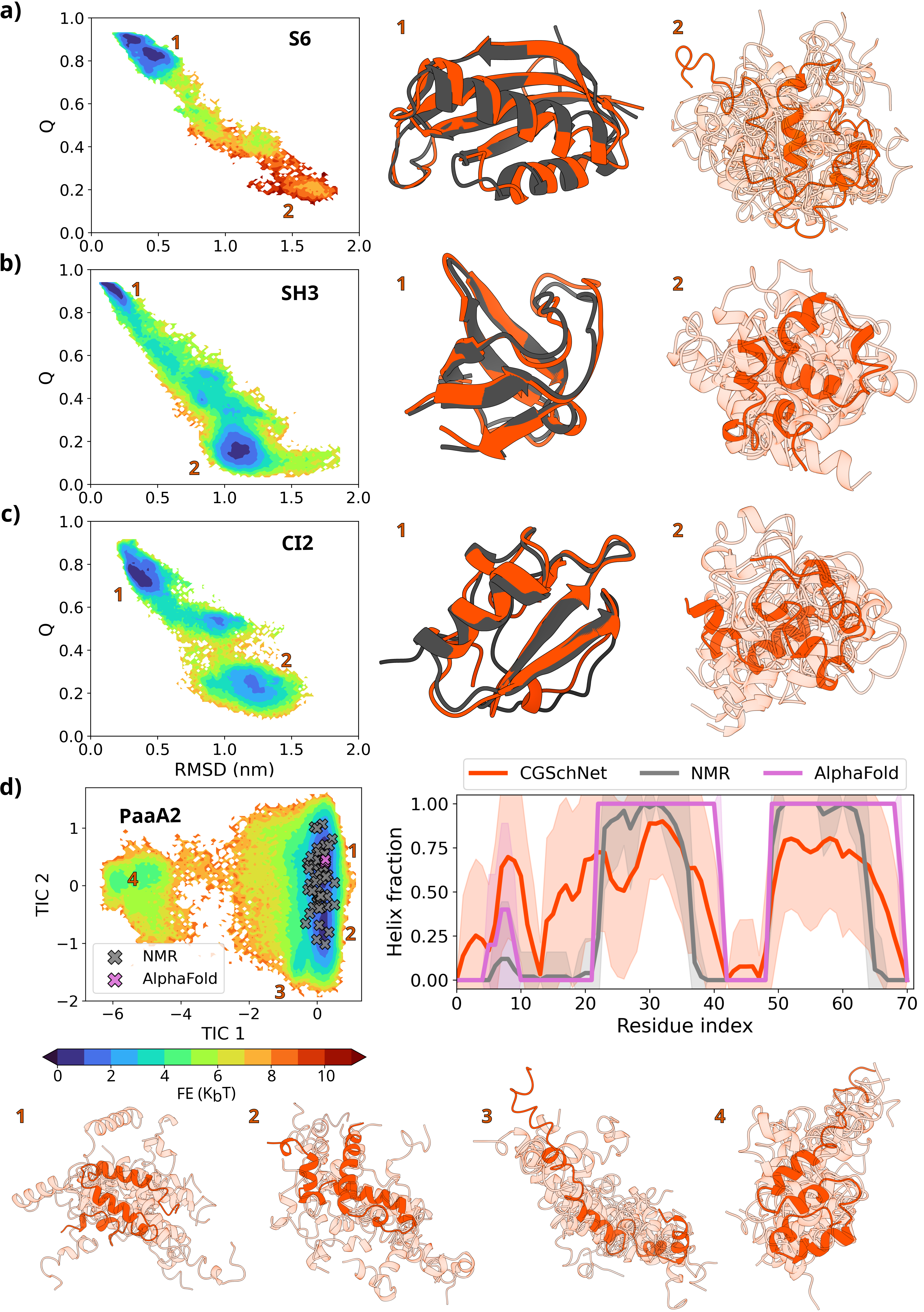}
\caption{Performance of the transferable CGSchNet on four large proteins outside of the training and validation sets for which there is no reference all-atom simulation data: 
a) 97-residue ribosomal protein S6 (1RIS), b) 55-residue SH3 domain (2NUZ), c) 65-residue CI-2 (2CI2) and d) Intrinsically disordered 71-residue antitoxin peptide PaaA2 (3ZBE). a-c, left: 2D free energy landscape as a function of the fraction of native contacts, $Q$, and the carbon alpha RMSD to the native state, obtained through PT simulations. Middle: structures from the most folded-like basin for CGSchNet (orange) and atomistic (grey) models. Right: illustrative structures extracted from the metastable basins indicated by the label. d, left: Free energy surface of PaaA2 as a function of the two first TICA components~\cite{Perez-Hernandez_2013}, obtained through Langevin simulations, with structures from the NMR ensemble and from the AlphaFold prediction marked with an x. Right: helix fraction per residue resulting from the DSSP~\cite{kabsch1983dictionary} analysis on our CG simulations, the NMR ensemble, and the AlphaFold prediction. Bottom: Representative structures extracted from different metastable states. 
\label{fig:large_proteins}}
\end{figure}

As an additional test, we also examine the extrapolative performance of our CG model on larger proteins for which we only have experimental reference data: three proteins with known PDB structure, and one intrinsically disordered protein with known helix fraction content from an ensemble of NMR structures.
The results are illustrated in Fig. \ref{fig:large_proteins}.
Specifically, we examine the native state stability of the 97-residue ribosomal protein S6~\cite{Lindahl_Svensson_Liljas_Sedelnikova_Eliseikina_Fomenkova_Nevskaya_Nikonov_Garber_Muranova_1994}, a 55-residue SH3 domain, and 65-residue CI-2 domain. These proteins have been routinely used as test proteins in protein folding studies in the past, both experimentally~\cite{itzhaki1995structure,martinez1998obligatory,otzen2002conformational,lindberg2007malleability} 
and computationally~\cite{li1994characterization,riddle1999experiment,Clementi_Nymeyer_Onuchic_2000,hubner2004simulation}.
Additionally, we simulate the conformational heterogeneity of the partially disordered 71-residue antitoxin peptide PaaA2~\cite{Sterckx_Volkov_Vranken_Kragelj_Jensen_Buts_Garcia-Pino_Jove_Van_Melderen_Blackledge_et_al_2014}.

For the structured proteins, we see that the CG simulations are capable of stabilizing structures near the folded state (Fig. \ref{fig:large_proteins}a-c). The CG model is also capable of sampling alternative states. For S6, a state is visited in which the sheet corresponding to residues 36 and 44 undergoes a secondary structure change to a helical segment, while the tertiary structure of the protein is maintained (Fig. \ref{fig:large_proteins}a). 

In the case of PaaA2, there is not a unique folded structure for comparison, but rather an NMR ensemble of 50 structures, wherein only two short helical domains are formed, while the rest part of the protein remains disordered. We see that the CG model is indeed qualitatively predicting the equilibrium helix fraction per residue in comparison to the NMR ensemble, although it somewhat overstructures the N-terminus (Fig. \ref{fig:large_proteins}d). 

\subsection{Detailed analysis and quantitative comparisons with other CG force fields}
\begin{figure}[ht]
    \centering    \includegraphics[width=\textwidth]{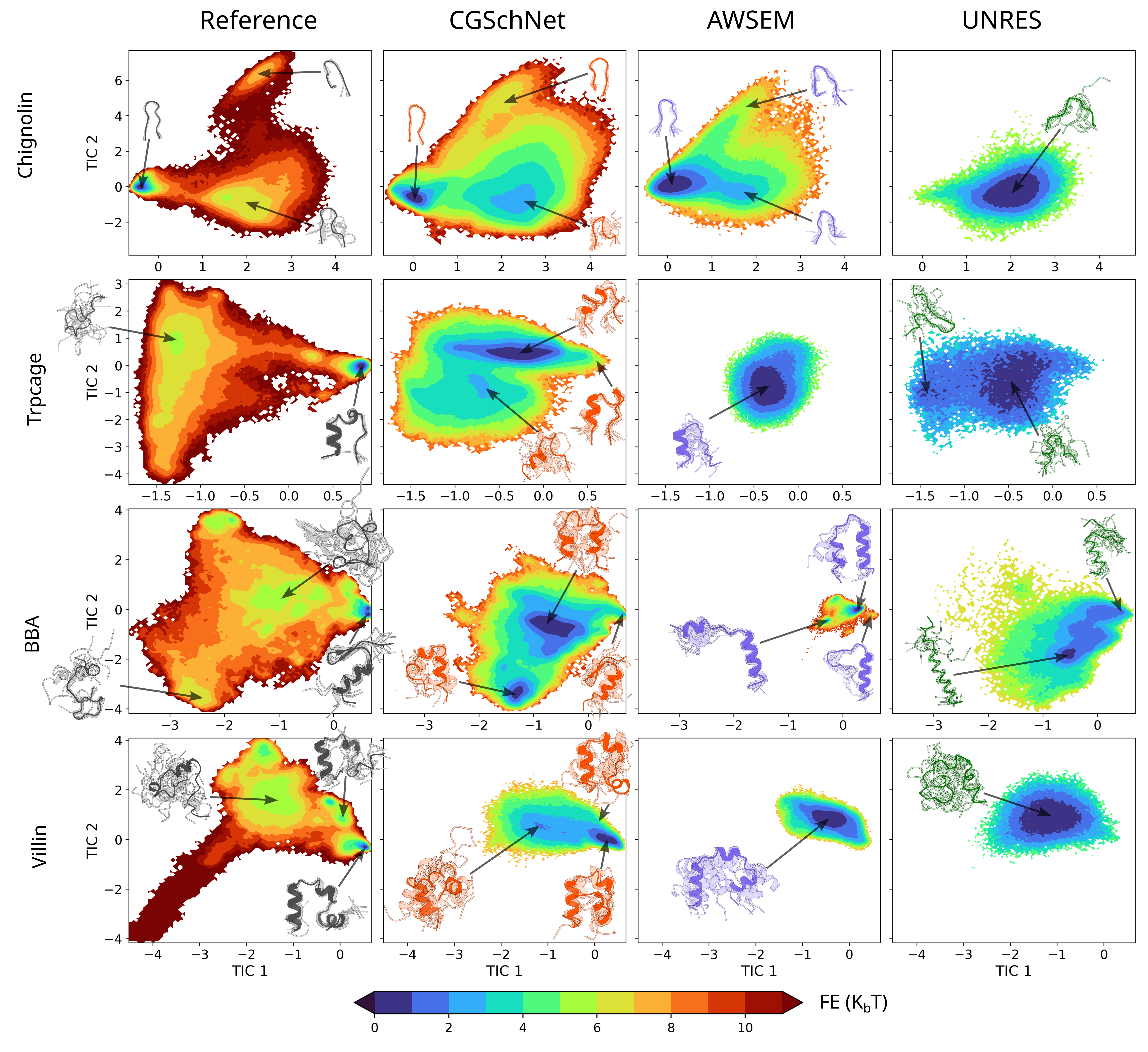}
\caption{Free energy landscapes as a function of the first two TICA coordinates~\cite{Perez-Hernandez_2013} for the four small proteins shown in Fig. \ref{small_protein_results}, for the reference atomistic simulations, and Langevin simulations of our CG model (CGSchNet) and other two different CG force fields of a comparable
resolution, AWSEM and UNRES. 
On each landscape representative structures from different metastable minima are shown. See SI for details on how the minima and the structures are selected.
\label{fig:comparison_landscapes}}
\end{figure}

Finally, we take a more detailed view of the characteristics of the learned CG energy landscapes and compare them with two other CG force fields with similar resolution, AWSEM (top-down)~\cite{Davtyan_Schafer_Zheng_Clementi_Wolynes_Papoian_2012} 
and UNRES (bottom-up/top-down)~\cite{Liwo_Sieradzan_Lipska_Czaplewski_Joung_Zmudzinska_Halabis_Oldziej_2019} (see SI for details on AWSEM and UNRES simulations). It is important to note that AWSEM, as a top-down CG force field, is 
parametrized to stabilize native states~\cite{Davtyan_Schafer_Zheng_Clementi_Wolynes_Papoian_2012}, which for all extrapolative targets presented should be stable at this temperature. Similarly, UNRES is parametrized with conformation data for multiple systems at several temperatures at approximately 300K~\cite{Liwo_Sieradzan_Lipska_Czaplewski_Joung_Zmudzinska_Halabis_Oldziej_2019}.

Fig. \ref{fig:comparison_landscapes} 
shows the free energy landscapes of the four small fast-folding proteins from Fig. \ref{small_protein_results} as a function of the two slowest TICA coordinates. For these four proteins, extensive MD simulations were done using the reference all-atom model, CGSchNet, AWSEM, and UNRES. We first note that the all-atom landscape exhibits the most structure and has the most metastable states, whereas the CG landscapes are smoother. CGSchNet captures the general form of the all-atom free energy landscape best and it also clearly resolves folded and unfolded states as well as separate metastable states, which is only observed in a few instances with AWSEM and UNRES. In many cases, AWSEM and UNRES only seem to have a single metastable state which is either folded or unfolded. We note that both force fields have been primarily parametrized for stabilizing proteins with a more pronounced fold, and this different objective is apparent when comparing whole free energy landscapes of less stable proteins.
Interestingly there is also a lot of resemblance between the all-atom reference and our machine-learned CGSchNet beyond the folded state prediction. In Chignolin, all three all-atom main states (folded, misfolded, and unfolded) are present in CGSchNet and AWSEM, but in CGSchNet these are also clearly metastable, whereas AWSEM is more down-hill. UNRES does not fold Chignolin. In Trpcage, BBA, and Villin the differences between ASWEM, UNRES, and the all-atom reference are even more striking as CGSchnet captures several of the metastable states of the all-atom reference, in particular those with a partial fold. However, there are significant differences in the unstructured states between the all-atom reference and all CG models, including ours, indicating that there is still room for improvement.

\begin{figure}[ht]
    \centering    \includegraphics[width=\textwidth]{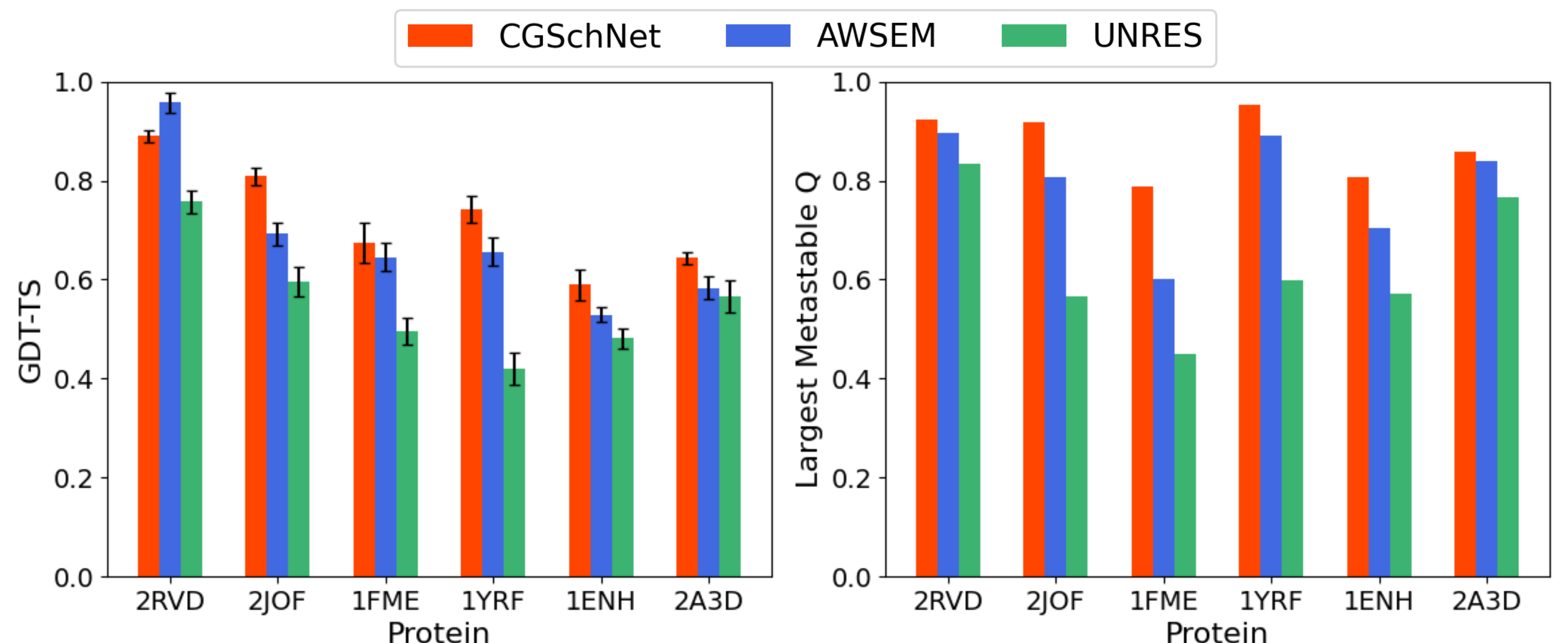}
\caption{Comparisons of transferable CGSchNet against two other CG force fields of a comparable resolution, AWSEM and UNRES, as assessed by the ability to stabilize structures near known native states. None of the proteins shown were used to parametrize the transferable CGSchNet. a) Average GDT total scores for several proteins over 10 random structures sampled from the most folded-like metastable state on the 2D Q vs. RMSD free energy surface for each CG force field (details in the SI). Error bars correspond to standard deviations in the scores across the 10 structures. b) Highest local maxima recovered from the 1D $Q$ probability distribution for each CG force field over all proteins.
\label{cg_ff_comparison}}
\end{figure}

We finally make a quantitative comparison between the all-atom model and the three CG force fields. For this, we focus on their ability to stabilize and find structure in the folded state. We sample 10 random structures from the metastable minimum on the free energy landscape with the largest $Q$/lowest RMSD and compute the average GDT-TS scores~\cite{Zemla_Venclovas_Moult_Fidelis_1999} with respect to the corresponding reference folded structure (see the SI for details). The GDT-TS score, introduced as a standard metric of quality of structure prediction in CASP~\cite{Zemla_Venclovas_Moult_Fidelis_1999}, measures the deviation of a predicted structure from a known structure in terms of the maximum percentage of C$_\alpha$ atoms which can be aligned (see SI). As a second metric, we record the 
$Q$ values corresponding to these same metastable states for direct comparison between the different CG models. Both metrics assess the accuracy of structure prediction and allow direct comparisons between different CG models.

Fig.~\ref{cg_ff_comparison} summarizes the results of these comparisons. Our transferable CGSchNet model predicts metastable states with comparable or lower RMSD/higher $Q$ for all the target proteins compared to AWSEM or UNRES. For all proteins, transferable CGSchNet also stabilizes structures with comparable or higher GDT total scores on average. Remarkably, the problem in the stabilization of native-like states for proteins with significant beta-sheet content, such as BBA, is reflected also in the other CG force fields. Still, both transferable CGSchNet and AWSEM are able to stabilize a near-native state of BBA accurately. This may suggest that BBA is a challenging target for transferable extrapolation in general. 
It is interesting to measure these metrics also on the results of atomistic simulations from D.E. Shaw research group~\cite{larsen_fast_folder_2011}.
In Supplementary Fig. S8, we show that even for these atomistic simulations the GDT-TS score is not always close to 1. In particular, in the case of Homeodomain (1ENH), the folded-like metastable state is at a Q-value around 0.6, lower than in our CG model.

\section{Discussion and Conclusion}
We have shown that it is possible to machine-learn a transferable, bottom-up, coarse-grained effective force field, that can be used for successful simulation on target proteins possessing little sequence overlap with proteins used for model parametrization. Notably, the training data is extremely small - 50 small protein domains and 1245 dimers of mono- or dipeptides. 
We have demonstrated with several examples that the model samples similar conformational spaces as an explicit water all-atom model, but orders of magnitude faster (see SI for the evaluation of the speedup, table S11). 
Indeed the model can be used to characterize the folding/unfolding free energy landscape of larger proteins that are unaffordable with atomistic MD. This significant improvement in speed was achieved using an in-house CG MD code that has not been optimized for efficient simulation. This increased performance is particularly notable as the code used for atomistic simulations is highly refined and MD simulations with neural network potentials are typically still much slower than those with classical force-fields.

The key property of the CG model that enables this performance is the fact that the effective CG energy is represented by a deep Graph Neural Network, capturing multi-body terms without imposing restrictive low-dimensional functional forms. The importance of multi-body terms in CG models has been extensively discussed in the literature~\cite{Ben-Naim_1997,Vendruscolo_Domany_1998,ejtehadi2004three,wang2009comparative,molinero2009water,larini2010multiscale,zimmermann2011free,gniewek2011multibody,larini2012coarse,das2012multiscale,john2017many,scherer2018understanding,Wang_Charron_Husic_Olsson_Noe_Clementi_2021,zaporozhets2023multibody}. Although it is expected that neural networks can capture the important multi-body effects, we believe it is remarkable that such a CG force field is \textit{transferable} in sequence space, especially given the rather small sequence coverage of the training set. A trade-off between structural accuracy and transferability has been observed in the past for a variety of CG protein models~\cite{tozzini2010minimalist,kar2014recent}.
However, CG effective energy functions have been previously parameterized with pre-designed functional forms, restricting the expressivity of multi-body energy terms and in practice precluding the possibility of quantitatively investigating the accuracy/transferability trade-off in protein systems. A deep neural network is the natural answer to such a problem and allows us to address this challenge. This is not the first instance of a bottom-up machine learning-based protein CG model~\cite{lemke2017neural,Wang_2019,Husic_Charron_Lemm_2020,Wang_Charron_Husic_Olsson_Noe_Clementi_2021,Majewski_Perez_Tholke_Doerr_Charron_Giorgino_Husic_Clementi_Noe_DeFabritiis_2022,Ding_Zhang_2022,Koehler_Chen_Kraemer_Clementi_Noe_2023,chennakesavalu2023ensuring,kramer_noise_opt_2023,wellawatte2023neural,airas2023transferable,ArtsEtAl_TwoForOne} but previously proposed versions were not transferable to proteins significantly different than those used for training.

The model presented in this work is designed with a bottom-up approach, which means, it aims to reproduce the properties of an atomistic model, and the variational force-matching approach is used for its optimization (see SI for details). 
The particular neural network chosen here (SchNet~\cite{Schutt_Sauceda_Kindermans_Tkatchenko_Muller_2018}) is quite simple and by no means the latest architecture for molecules. It consists of a series of continuous-filter convolutions and does not include an attention mechanism, nor an explicit long-range interaction term. 
This architectural choice was motivated by the goal of developing a ``proof of concept'' model, that can be trained and simulated as fast as possible while still yielding the desired results. We believe that more sophisticated architectural choices could produce better-performing CG models. In particular, the lack of long-range interactions in our model may affect the model performance on much larger and multi-protein systems, where electrostatic interactions may play an important role~\cite{sheinerman2002role,zhang2011role}. Multiple approaches for including long-range interactions~\cite{ko2021fourth,Unke_Chmiela_Gastegger_Schutt_Sauceda_Muller_2021,frank2022so3krates,kosmala2023ewald} or attention mechanisms~\cite{velivckovic2017graph,tholke2021equivariant,han2022geometrically} have been recently proposed and could be incorporated into our modeling framework.

It is important to note that our CG model was trained at a specific thermodynamic condition. Transferability in temperature/pressure or other additional environmental parameters is therefore not expected at this point. In particular, the temperature dependence of the CG effective energy is highly nontrivial, as it is in fact a free energy with an entropic component~\cite{dunn2016van,kidder2021energetic,jin2019understanding}. An explicit dependence of the model on thermodynamic conditions could be in principle included in the framework~\cite{krishna2009multiscale,izvekov2010multiscale,das2010multiscale,rosenberger2018addressing,lebold2019dual,ruza2020temperature,pretti2021microcanonical}. However, in practice, its training will likely require the curation of a significantly larger dataset encompassing multiple simulations at multiple thermodynamic conditions, which will likely require large-scale computational resources.

The results were obtained with a model that, while aggressively coarse-grained with respect to an explicit water atomistic model, still retains the full backbone heavy atoms and an additional atom per side-chain (excluding Glycine). We have at this point not investigated alternative resolutions. We expect the model transferability to be strongly tied to the model resolution. Although different methods have been proposed for the simultaneous optimization of CG effective energy and CG mapping~\cite{schoberl2017predictive,wang2019coarse,chennakesavalu2023ensuring}, it remains unclear if and which additional resolutions allow the design of a transferable and quantitatively accurate model. We believe the results presented here open the way to a systematic investigation of this point.

\section{Methods}
\subsection{Neural network model}
Building on previous efforts~\cite{Husic_Charron_Lemm_2020, Chen_Kramer_Charron_Husic_Clementi_Noe_2021, kramer_noise_opt_2023}, we choose to model the optimizable term of our CG effective energy with the graph neural network architecture, CGSchNet, which is based on a previous architecture, SchNet~\cite{Schutt_Sauceda_Kindermans_Tkatchenko_Muller_2018}. See SI for a detailed description of network architecture, hyperparameter choices, and training routines. The ability to directly learn species-dependent interactions and CG bead-wise features from data represents the primary advantage of using a convolutional graph neural network such as CGSchNet in this pursuit. More recent graph neural network architectures~\cite{Batzner_Musaelian_Sun_Geiger_Mailoa_Kornbluth_Molinari_Smidt_Kozinsky_2022,Batatia_Kovacs_Simm_Ortner_Csanyi_2022} may be used alternatively. However, we note that newer architectures often require more computational resources, which may create significant barriers given the large number of MD simulation frames and the system sizes used in training.

\subsection{Loss function}
We design our CG model within the framework of variational force-matching 
\cite{Izvekov_Voth_2005,Noid_Chu_Ayton_Krishna_Izvekov_Voth_Das_Andersen_2008}. In practice, we optimize the parameters $\{\theta\}$ of a network representing the effective energy $\tilde{U}_{CG}(\bold{R}; \{\theta\})$, where $\bold{R}$ are the CG coordinates, by minimizing a loss function in the form:
\begin{equation}
\chi^2 [\tilde{\bold{F}}_{CG,\Delta}(\bold{R};\{\theta\})] = \left \langle \frac{1}{3N} 
\sum_{j=1}^{N} \left\lVert \left[\mathcal{M}_F\bold{f}_{AA}(\bold{r})\right]_{\Delta,j} - \tilde{\bold{F}}_{CG, \Delta}(\bold{R}; \{\theta\})_j \right\rVert^2 \right \rangle_{\bold{r}}
\label{delta_force_matching_functional}
\end{equation}

Here, $N$ is the number of CG atoms in the system. In the equation above, $\tilde{\bold{F}}_{CG, \Delta}(\bold{R}; \{\theta\})$
are the forces associated with the CG effective energy, $\tilde{\bold{F}}_{CG}(\bold{R}; \{\theta\}) = -\nabla_{\bold{R}}\tilde{U}_{CG}(\bold{R}; \{\theta\})$, after subtraction of the ``prior forces'':
\begin{equation}
\tilde{\bold{F}}_{CG, \Delta}(\bold{R}; \{\theta\}) = \tilde{\bold{F}}_{CG}(\bold{R}; \{\theta\}) - \bold{F}_\text{prior}(\bold{R}),
\label{delta_decomp}
\end{equation}
where
$\bold{F}_\text{prior}(\bold{R}) = - \nabla_{\bold{R}}\tilde{U}_\text{prior}(\bold{R})$ and $\tilde{U}_\text{prior}(\bold{R})$ is a pre-fit ``prior energy'' term. The atomistic force is similarly modified. The definition of a prior energy is discussed in the next section and it has been shown to play an important role in constructing stable and accurate neural network-based CG models by enforcing asymptotic physical behavior in regions of phase space not covered adequately by a training dataset~\cite{Wang_2019, Husic_Charron_Lemm_2020, Chen_Kramer_Charron_Husic_Clementi_Noe_2021, Durumeric_Charron_Templeton_Musil_Bonneau_Pasos-Trejo_Chen_Kelkar_Noe_Clementi_2023}. 
In Eq.~\ref{delta_force_matching_functional}, the operator $\mathcal{M}_F$ projects the atomistic forces $\bold{f}_{AA}(\bold{r})$, as a function of the atomistic coordinates $\bold{r}$, on the CG space. We have shown in previous work that a careful choice of $\mathcal{M}_F$ is crucial to the optimization of the CG model~\cite{kramer_noise_opt_2023}. In this work, $\mathcal{M}_F$ is defined for each CG site as the sum of forces on the preserved atom and neighboring hydrogen atoms connected via constrained bonds~\cite{kramer_noise_opt_2023}.

\subsection{CG resolution and prior energy}
A good choice of the prior energy model should be connected to the resolution chosen to define the CG model~\cite{Durumeric_Charron_Templeton_Musil_Bonneau_Pasos-Trejo_Chen_Kelkar_Noe_Clementi_2023}. Previous non-transferable CG studies~\cite{Husic_Charron_Lemm_2020, kramer_noise_opt_2023, Koehler_Chen_Kraemer_Clementi_Noe_2023} have utilized a resolution that retains only the carbon alpha ($C_\alpha$) atoms for each amino acid. However, when considering 20 naturally occurring amino acids, the type enumeration for common local energy terms, such as dihedral interactions, becomes very large. While efforts in the past~\cite{Jr_Lu_Voth_2010} have attempted to mitigate such scaling, this can lead to potentially limiting or overly-biasing expressions for the associated prior energies.

For this work, we choose to retain the following 5 atoms for each residue (4 for GLY residues, which do not comprise a $C_\beta$): $N$, $C_\alpha$, $C_\beta$, $C$, and $O$. This 5-bead-per-residue CG mapping is not unprecedented; the successful AWSEM~\cite{Davtyan_Schafer_Zheng_Clementi_Wolynes_Papoian_2012} CG force field, which retains the $C_\alpha$, the $C_\beta$, and the $O$ atoms (as well as virtual sites for $N$ and $C$ atoms), utilizes a comparable resolution.

This choice of CG resolution allows for a direct interpretation of secondary structures and leads to intuitive prior energy choices (e.g., physical bond/angle terms, physical dihedral angles, etc.)  For a description of the terms in the prior force field, we refer the reader to the SI. 

It is important to stress that if the prior energy is used alone (without the trainable neural network energy term $\tilde{U}_{CG}(\bold{R}; \{\theta\}$) it is completely incapable of stabilizing any secondary or tertiary protein structures (see SI for the results from control simulations). The function of the prior energy is only to prevent the model from visiting configurational unphysical regions (e.g. overlapping atoms), with no additional bias on the configurational landscape (see SI). 

\subsection{Training data}
In order to capture sequence and secondary/tertiary structure diversity, we construct a dataset of native state all-atom simulations of 50 protein domains from the CATH~\cite{cath} database (see SI for the domain selection procedure). Each simulation represents $100,000$ frames of MD data wherein solute forces and positions are saved. In addition to this dataset of folded CATH simulations, we also construct a second dataset wherein $\approx$1200 mono/dipeptide dimer systems are simulated using umbrella sampling with dimer center of mass distances as a reaction coordinate, each system consisting of $27,000$ frames. While this dataset contains no direct secondary/tertiary structure information, it contains valuable, asymptotic force information for atoms that are brought very close together through the nature of the sampling strategy.

A comprehensive discussion about the training and validation datasets that are used to parameterize the model can be found in the SI.

\section*{Supplementary information}
Supplementary text and figures are available with details on all the methodologies used. 

\section*{Data availability}
Simulation data required for reproduction of the results shown in this manuscript will be made public upon article publication.

\section*{Code availability}
Codes required for reproduction of the results associated with this manuscript will be made public upon article publication.

\section*{Acknowledgements}
We thank all members of the Clementi and No\'e groups for their help in different phases of this work. 
We gratefully
acknowledge funding from the European Commission (Grant No.~ERC CoG 772230
\textquotedblleft ScaleCell\textquotedblright), the International
Max Planck Research School for Biology and Computation (IMPRS\textendash BAC),
the BMBF (Berlin Institute for Learning and Data, BIFOLD), the Berlin
Mathematics center MATH+~(AA1-6, EF1-2) and the Deutsche Forschungsgemeinschaft
DFG (NO 825/2, NO 825/3, NO 825/4, GRK DAEDALUS, SFB/TRR 186, Project A12; SFB 1114, Projects B03, B08, and A04; SFB 1078, Project C7; and RTG 2433, Project Q05, Q04), the National Science Foundation (CHE-1900374, and PHY-2019745), and the Einstein Foundation Berlin (Project 0420815101). The authors gratefully acknowledge the computing time provided on the supercomputer Lise at NHR@ZIB as part of the NHR infrastructure. The authors thank volunteers at GPUGRID.net for contributing computational resources and Acellera for funding.

\input{MLCG.bbl}

\end{document}